# Theoretical Investigation on the Effect of multinary Isoelectronic Substitution on TiCoSb based half-Heusler alloys

Mukesh K. Choudhary[1, 2] and P. Ravindran[1, 2, a)]

[1]*Department of Physics, Central University of Tamil Nadu, Thiruvarur-610005*
[2]*Simulation Center for Atomic and Nanoscale Materials, Central University of Tamil Nadu, Thiruvarur-610005*
[a)] Email: raviphy@cutn.ac.in

**Abstract.** To understand the effect of isoelectronic substitution on thermoelectric properties of TiCoSb based half - Heusler (HH) alloys, we have systematically studied the transport properties with substitution of Zr at Ti and Bi at Sb sites. The electronic structure of $Ti_xZr_{1-x}CoSb_xBi_{1-x}$ ($x$ = 0.25, 0.5, 0.75) and parent TiCoSb are investigated using the full potential linearized augmented plane wave method and the thermoelectric transport properties are calculated on the basis of semi-classical Boltzmann transport theory. The band analysis of the calculated band structures reveal that $Ti_xZr_{1-x}CoSb_xBi_{1-x}$ has semiconducting behavior with indirect band gap at $x$ = 0.25, 0.5 concentration and direct band gap behavior at $x$ = 0.75 concentration. The $Ti_xZr_{1-x}CoSb_xBi_{1-x}$ ($x$ = 0.25, 0.5, 0.75) compounds show smaller band gap values as compared to the pure TiCoSb. The d electrons of Ti/Zr and Co dominate the electronic transport properties of $Ti_xZr_{1-x}CoSb_xBi_{1-x}$ system. All these systems follow the empirical rule of 18 valence-electron content to bring semiconductivity in HH alloys. The isoelectronic substitution in TiCoSb can tune the band structure by shifting the Fermi level. This provides us lot of possibilities to get the desired band gap values for designing thermoelectrics with high efficiency. In this study we have showed that the isoelectronic substitution at both Ti and Sb site of TiCoSb has very small effect for increasing the *ZT* values and one should go for isoelectronic substitution at any one sites of TiCoSb HH alloys alone to improve *ZT*.

## INTRODUCTION

With the rise in economy in the 21st century world has urge more demand of the energy consumption for everyday work which produces the harmful chemical and gases to the environment and hence increase the global warming. In the present scenario there has been lots of research developed to find the alternative energy technologies[1] to reduce the energy supply gap. Thermoelectric materials (TE) can be the alternative solution to reduce the energy consumption by converting the waste heat energy into useful electrical energy. One of the major challenge for the researchers is to find the low cost, light weight and environment friendly materials for waste heat recovery. HH alloys are considered as the good candidates to deliver all the requirements. HH alloys are ternary intermetallic compounds with the general formula *XYZ* in which *X* and *Y* are typically transition metals and *Z* is a main group element. The efficiency of the TE materials can be calculated as the dimensionless figure of merit $ZT = S^2\sigma T/\kappa$, where S is Seebeck coefficient, $\sigma$ is electrical conductivity, $\kappa$ is total thermal conductivity i.e. electronic and thermal conductivity ($\kappa_e + \kappa_l$)[2, 3]. HH alloys conserve the semiconducting behavior by transferring the valence electrons from electropositive element *X* to more electronegative elements *Y* and *Z* provide the stable closed shell configuration i.e. a $d^{10}$ for *Y* and $s^2p^6$ configuration for *Z*. HH alloys based on TiCoSb has been investigated extensively over a past decades. Experimentally it has been found that TiCoSb has high power factor. But it also shows high thermal conductivity which decreases the efficiency. One of the successful approach adopted by the researchers to reduce the thermal conductivity is isoelectronic substitutions of different size atom at the sites of Ti, Co and Sb of TiCoSb alloys[4]. Isoelectronic substitutions cause additional phonon scattering centre which reduces the thermal conductivity and hence expected to increase the efficiency of the system. In our previous[5] studies we have reported that the multinary substitution without altering the 18 valance electron count rule (VEC) can increase the phonon scattering center in the system and hence increases the *ZT*. In the present study we have focused on calculating the effect of isoelectronic substitution on both Ti and Sb sites of TiCoSb alloys.

# COMPUTATIONAL DETAILS

The structural optimization was made using the Vienna *Ab-initio* simulation package (VASP)[6,7] within the projected augmented plane wave (PAW) method to obtained the ground state structure. The Perdew-Burke-Ernzerhof generalized gradient approximation (GGA)[8] is used for the exchange correlation potential. We have used the supercell approach for the isoelectronic substitution in TiCoSb. The energy convergence criterion was chosen to be $10^{-6}$ eV and the cut-off energy of the plane wave was set to be 600 eV, and the pressure on the cell had minimized within the constraint of constant volume. A 12×12×12 **k**-mesh was used for the structure optimization. A full-potential augmented plane wave method as implemented in the Wien2k code with the very large 31×31×31 **k**-mesh has been used to calculate the accurate electronic structure for calculating the thermoelectric transport properties. The electrical transport properties such as Seebeck coefficient, electrical conductivity, electronic part of thermal conductivity, and thermoelectric *ZT* were calculated by using the BoltzTraP[9] code.

**TABLE 1.** Optimized lattice constants (a), band gap values (Eg), and thermoelectric figure of merit ZT for TixZr1-xCoSbxBi1-x where IB and DB are indirect and direct band gap respectively.

| x | a(Å) | $E_g$ (eV) | ZT 300K | ZT 700K |
|---|---|---|---|---|
| 0.25 | 6.14 | 0.94 (IB) | 0.97 | 0.95 |
| 0.50 | 6.08 | 0.93 (IB) | 0.97 | 0.95 |
| 0.75 | 5.97 | 1.01 (DB) | 0.98 | 0.96 |
| 1.00 | 5.87 | 1.04 (IB) | 0.98 | 0.96 |
| $Ti_{0.5}Zr_{0.25}Hf_{0.25}CoPb_{0.5}Te_{0.5}$ (Previous work) | 6.03 | 0.93 (DB) | 1.05 | 0.97 |

# RESULTS AND DISCUSSION

The crystal structure of HH alloys are shown in Fig. 1. These compounds crystallize in the cubic structure with the space group $F\bar{4}3m$ (No.216) and shows the large variety of physical properties to obtain the high efficient TE materials. The ground state structural parameters were calculated by relaxation of both lattice parameter and atomic positions Table 1 summarized the optimized lattice parameters, calculated band gap values and *ZT* of $Ti_xZr_{1-x}CoSb_xBi_{1-x}$ system.

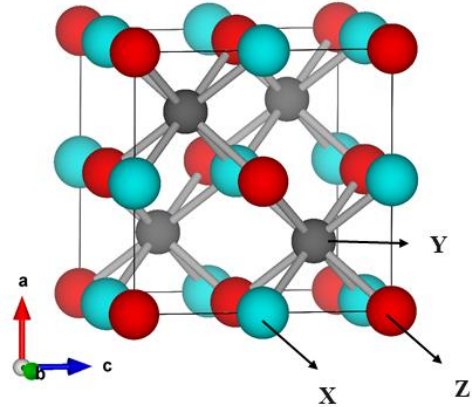

**FIGURE 1.** Crystal structure of half- Heusler compound.

For simulating the $Ti_xZr_{1-x}CoSb_xBi_{1-x}$ we have modeled the supercell by varying the concentration of 25, 50 and 75% of Zr at Ti site and with same concentration of Bi at Sb site. The calculated lattice constants for $Ti_xZr_{1-x}CoSb_xBi_{1-x}$ ($x$ = 0.25, 0.5, 0.75) are decreasing with $x$ as given in Table 1. We have calculated the orbital projected energy band structure of $Ti_xZr_{1-x}CoSb_xBi_{1-x}$ to understand the orbital contributions to bands at the band edges. Figure 2 shows the orbital projected electronic band structure of $Ti_xZr_{1-x}CoSb_xBi_{1-x}$ as a function of $x$ closer to their band edges i.e. from -3 eV to 3 eV. From the Fig. 2(a) we can see that the valence band maximum (VBM) and conduction band minima (CBM) of pure TiCoSb are mainly occupied by Co $d_{xz}$, $d_{x^2-y^2}$, $d_{z^2}$ and Ti/Zr $d_{xy}$, $d_{yz}$ states respectively with small contribution of Sb/Bi $p_x$, $p_y$, $p_z$ at -3 eV and 3 eV. Similar trend has also been observed in $Ti_xZr_{1-x}CoSb_xBi_{1-x}$ systems. All these alloys exhibit finite bandgap. The compounds TiCoSb, $Ti_{0.25}Zr_{0.75}CoSb_{0.25}Bi_{0.75}$, and $Ti_{0.5}Zr_{0.5}CoSb_{0.5}Bi_{0.5}$, exhibit indirect band gap behavior with VBM at Γ, L, M and the CBM at X, Γ, R point, respectively. In the case of $Ti_{0.75}Zr_{0.25}CoSb_{0.75}Bi_{0.25}$ show direct band gap behavior with both VBM and CBM at the Γ point in the irreducible wedge of the first Brillouin zone of the simple cubic lattice. Our calculated band gap for TiCoSb is 1.04 eV which is in good agreement with previous report[10], while substituted $Ti_xZr_{1-x}CoSb_xBi_{1-x}$ systems shows the smaller band gap values than that in TiCoSb (see the Table 1). Our orbital projected band structure reveal that in HH alloys the d state plays an important role for the electronic transport properties.

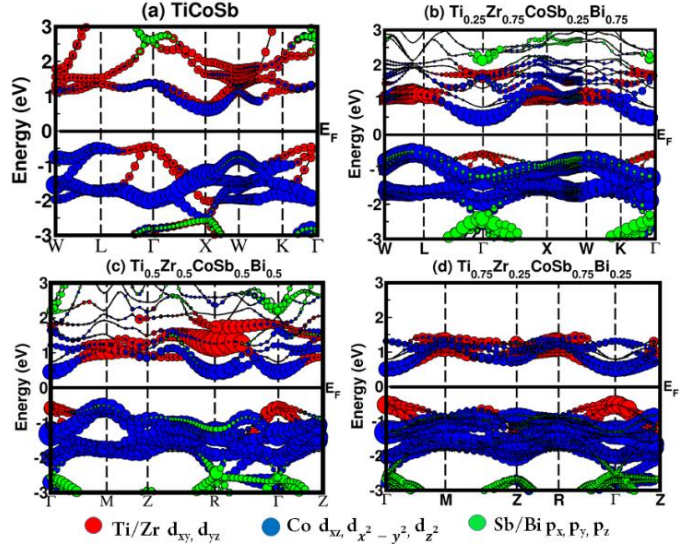

**Figure 2**. Band structure of (a) TiCoSb (b) $Ti_{0.25}Zr_{0.75}CoSb_{0.25}Bi_{0.75}$ (c) $Ti_{0.5}Zr_{0.5}CoSb_{0.5}Bi_{0.5}$ (d) $Ti_{0.75}Zr_{0.25}CoSb_{0.75}Bi_{0.25}$.

The calculated electronic structures have been used to calculate the thermoelectric transport properties of $Ti_xZr_{1-x}CoSb_xBi_{1-x}$ Figure 3 shows the transport properties i.e. S, $\kappa_e$, and dimensionless thermoelectric ZT as a function of chemical potential for $Ti_xZr_{1-x}CoSb_xBi_{1-x}$ in the range of -1 eV to 1 eV at 300K and 700K. The S of $Ti_{0.5}Zr_{0.5}CoSb_{0.5}Bi_{0.5}$ has the lower value compared to other considered systems (see the Fig. 3(a) and 3(b)). Further S of $Ti_xZr_{1-x}CoSb_xBi_{1-x}$ is almost double at room temperature compared with that at 700K.

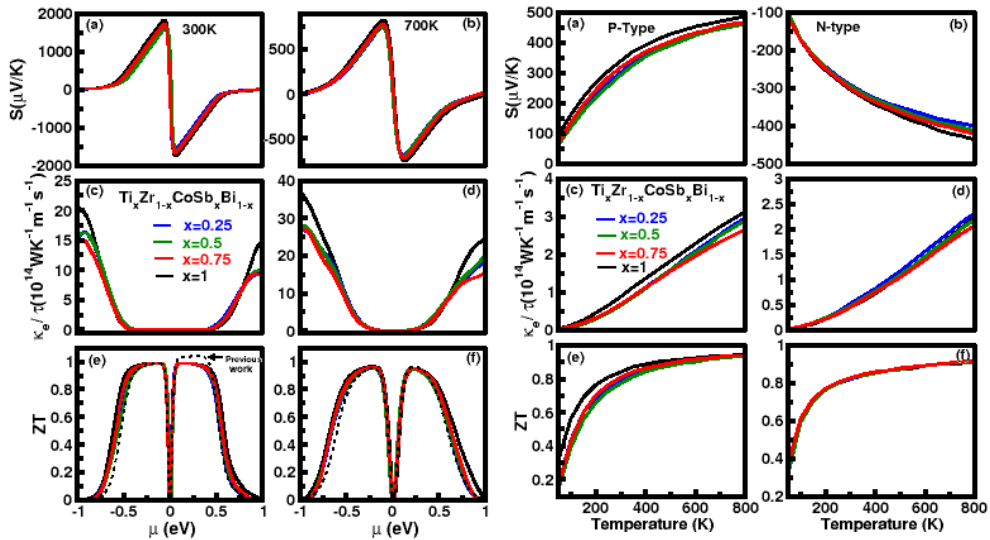

**FIGURE 3.** Transport properties of $Ti_xZr_{1-x}CoSb_xBi_{1-x}$ (x = 0.25, 0.5, 0.75, 1) as a function of chemical potential at 300K and 700K (left panel) and as a function of temperature (right panel) respectively. Here black dotted line show the ZT for $Ti_{0.5}Zr_{0.25}Hf_{0.25}CoPb_{0.5}Te_{0.5}$ (previous work) as a function of chemical potential at 300K and 700K for comparison.

We have calculated the dimensionless thermoelectric *ZT* from the Seebeck coefficient (S) electrical conductivity ($\sigma/\tau$) and the electronic part of thermal conductivity ($\kappa_e/\tau$) as a function of chemical potential and temperature. We have found that the electronic thermal conductivity of $Ti_xZr_{1-x}CoSb_xBi_{1-x}$ shows the smaller value at $x = 0.75$ substitution (Fig 3 (c, d)) which indicates that for small concentration of Zr and Bi substitution at Ti and Sb sites introduce more scatering centre into the system and hence increase the *ZT* value. Also we are suggesting that the small concentration of bigger and heavy element can reduce the $\kappa_e$ and increase the efficiency. Furthermore our calculated *ZT* value as a function of chemical potential show very small increment at $x = 0.75$ substituted $Ti_xZr_{1-x}CoSb_xBi_{1-x}$ system. Here we want to underline our previous work (cited above) which shows that the Hf substituted $Ti_{0.5}Zr_{0.25}Hf_{0.25}CoPb_{0.5}Te_{0.5}$ with electron doping (n-type condition) achieved the high *ZT* value of 1.05 indicating that the extra Hf substitution can strongly influence the transport properties of this system. In addition, we have also calculated the S, $\kappa_e$ and *ZT* as a function of temperature for the hole doped and electron doped systems as shown in Fig. 3 (a-f right panel). Both p and n-type doping shows smaller S and $\kappa_e$ values compared with pure TiCoSb system. Furthermore the hole doped $Ti_xZr_{1-x}CoSb_xBi_{1-x}$ ($x = 0.25, 0.5, 0.75$) systems show smaller *ZT* value compared to electron doped systems those show little relatively higher ZT value and stay constant at higher temperature (see the right panel Fig. 3(e, f)).

## CONCLUSION

For the $Ti_xZr_{1-x}CoSb_xBi_{1-x}$ ($x = 0.25, 0.5, 0.75$) and TiCoSb systems we have performed the structural optimizations with VASP, electronic structure calculations with WIEN2k, and transport calculations with BoltzTraP codes. The thermoelectric related electrical transport properties of these systems, including S, $\kappa_e$ thermoelectric *ZT*, and their dependence on the Fermi levels, are investigated by combining the Boltzmann transport theory and the electronic structure obtained from WIEN2k. The orbital projected electronic band structures analysis shows that the d orbital of Ti/Zr and Co plays an important role in the transport properties of pure and isoeletronic substituted TiCoSb based HH alloys. Furthermore $Ti_xZr_{1-x}CoSb_xBi_{1-x}$ ($x = 0.25, 0.5, 0.75$,) systems show semiconducting behavior with smaller band gap values compared with pure TiCoSb. Also we found that the isoelectronic substitution at both Ti and Sb sites in TiCoSb has very small increase in the *ZT* value.

## ACKNOWLEDGEMENTS

The authors are grateful to the Department of Science and Technology, India for the funding support via Grant No. SR/NM/NS-1123/2013 and also supported by the Indo-Norwegian Cooperative Program (INCP) via Grant No. F. No. 58-12/2014(IC).

## REFERENCES

[1] J. Yang and F.R. Stabler, J. Electron. Mater. **38**, 1245 (2009).
[2] H. Geng, X. Meng, H. Zhang, J. Zhang, H. Geng, X. Meng, H. Zhang, and J. Zhang, Appl. Phys. Lett. **202104**, 1 (2016).
[3] G. Ding, J. Phys. D Appl. Phys (2014).
[4] P. Qiu, X. Huang, X. Chen, L. Chen, P. Qiu, X. Huang, X. Chen, and L. Chen, **103703**, (2009).
[5] M.K. Choudhary and P. Ravindran, 1 (2018).
[6] G. Kresse and J. Furthmiiller, Comput. Mater. Sci. **6**, 15 (1996).
[7] T. Physik, T.U. Wien, and W. Hauptstrasse, Phys. Rev. B **47**, (1993).
[8] J.P. Perdew, K. Burke, and M. Ernzerhof, Phys. Rev. Lett. 3865 (1996).
[9] G.K.H. Madsen and D.J. Singh, Comput. Phys. Commun. **175**, 67 (2006).
[10] L.L. Wang, L. Miao, Z.Y. Wang, W. Wei, R. Xiong, H.J. Liu, J. Shi, X.F. Tang, L.L. Wang, L. Miao, Z.Y. Wang, W. Wei, R. Xiong, H.J. Liu, and J. Shi, J. Appl. Phys. **13709**, (2011).